\documentclass[aps,prd,superscriptaddress,longbibliography,twocolumn,showkeys]{revtex4-2}
\usepackage{amssymb,amsmath}
\usepackage{verbatim}
\usepackage{graphicx}
\usepackage{xcolor}
\usepackage{hyperref}
\usepackage{orcidlink}
\usepackage{fontawesome}

\usepackage{bm}
\newcommand{\change}[1]{{\color{black}#1}}

\newcommand\Mpc{Mpc.$h^{-1}$ }
\newcommand\hbyMpc{$h\rm{Mpc}^{-1}$}
\newcommand{\github}{\href{https://github.com/nmudur/diffusion-hmc}{\faGithub}}
\begin{document}
\title{Diffusion-HMC: Parameter Inference with Diffusion-model-driven Hamiltonian Monte Carlo}

\author{Nayantara Mudur\,\orcidlink{0000-0001-5139-612X}}\email{nmudur@g.harvard.edu}
\affiliation{Department of Physics, Harvard University, 17 Oxford Street, Cambridge, MA 02138, USA}
\affiliation{Harvard-Smithsonian Center for Astrophysics, 60 Garden Street, Cambridge, MA 02138, USA}
\affiliation{The NSF AI Institute for Artificial Intelligence and Fundamental Interactions Massachusetts Institute of Technology, Cambridge, MA 02139, USA}

\author{Carolina Cuesta-Lazaro\,\orcidlink{0000-0002-6069-2999}}
\affiliation{Harvard-Smithsonian Center for Astrophysics, 60 Garden Street, Cambridge, MA 02138, USA}
\affiliation{The NSF AI Institute for Artificial Intelligence and Fundamental Interactions Massachusetts Institute of Technology, Cambridge, MA 02139, USA}
\affiliation{Department of Physics, Massachusetts Institute of Technology, Cambridge, MA 02139, USA}

\author{Douglas P. Finkbeiner\,\orcidlink{0000-0003-2808-275X}}
\affiliation{Department of Physics, Harvard University, 17 Oxford Street, Cambridge, MA 02138, USA}
\affiliation{Harvard-Smithsonian Center for Astrophysics, 60 Garden Street, Cambridge, MA 02138, USA}
\affiliation{The NSF AI Institute for Artificial Intelligence and Fundamental Interactions Massachusetts Institute of Technology, Cambridge, MA 02139, USA}

\begin{abstract}
Diffusion generative models have excelled at diverse image generation and reconstruction tasks across fields. A less explored avenue is their application to discriminative tasks involving regression or classification problems. The cornerstone of modern cosmology is the ability to generate predictions for observed astrophysical fields from theory and constrain physical models from observations using these predictions. This work uses a single diffusion generative model to address these interlinked objectives -- as a surrogate model or emulator for cold dark matter density fields conditional on input cosmological parameters, and as a parameter inference model that solves the inverse problem of constraining the cosmological parameters of an input field. The model is able to emulate fields with summary statistics consistent with those of the simulated target distribution. We then leverage the approximate likelihood of the diffusion generative model to derive tight constraints on cosmology by using the Hamiltonian Monte Carlo method to sample the posterior on cosmological parameters for a given test image. Finally, we demonstrate that this parameter inference approach is more robust to small perturbations of noise to the field than baseline parameter inference networks.
\end{abstract}

\maketitle

\section{Introduction}
Ongoing and upcoming missions, such as the Dark Energy Spectroscopic Instrument \href{https://www.desi.lbl.gov/the-desi-survey/}{DESI}, the Vera C. Rubin Observatory's \href{https://www.lsst.org/}{Legacy Survey of Space and Time} and the \href{https://roman.gsfc.nasa.gov/}{Nancy Grace Roman Space Telescope} will map the cosmos at unprecedented resolution and volume. This has created a proportionate demand for simulations that can generate predictions from theory. Cosmological simulations, however, are expensive to run, and can only be generated for a limited set of initial conditions and points in parameter space. 

The canonical summary statistic used for parameter inference is the two point correlation
function, or the power spectrum $Pk$ at large, linear scales where \change{perturbation theory holds. At smaller scales, however, gravitational collapse induces non-Gaussianity in the fields. This means the information content of cosmological fields is not fully captured by the power spectrum at the large scales alone.} For example, recent work \citep{SIMBIG,nguyen2024much} derived much stronger constraints
on $\sigma_8$ by going beyond the two point correlation function at linear scales and analyzing non-linear modes at smaller scales ($k >$0.25 \hbyMpc) or by extracting information at the field-level. A multitude of other statistics -- such as the marked power spectrum, the bispectrum, the wavelet scattering transform, and void probability functions -- have also been devised \citep{valogiannis2022going,paillas2023cosmological, regaldo2022generative,hamaus2016constraints} in an effort to capture higher order correlations in non-Gaussian fields. Previous work \citep{Heitmann:2009cu, mustafa2019cosmogan, paillas2023cosmological, valogiannis2023precise,sharma2024field} has addressed the prohibitive cost of simulations by creating emulators or surrogate models that learn to interpolate between predictions of a \textit{specific} summary statistic between training points using formalisms such as Gaussian processes. More recently, simulation-based inference at the field level has also been used, where a likelihood model parameterized by a neural network is learned on the fields \citep{dai2024multiscale,cuesta2023point}.

Generative models are a class of machine learning approaches that enable one to simulate the ability to draw samples from a complicated target probability density and include variational autoencoders, normalizing flows and generative adversarial networks (GANs). Diffusion generative models \citep{sohl2015deep,song2021scorebased} involve a forward diffusion (noising) process that transform samples from the target distribution to those from the standard normal. In the denoising diffusion probabilistic model (DDPM)\citep{ho2020denoising}, the noising process consists of a variance schedule ${\beta_t}$ over a fixed number of time steps, $T$, that determines the incremental noise added to the image. Since the diffusion process can be formulated as a stochastic differential equation (SDE) the DDPM variance schedule corresponds to the discretization of this SDE. In the generative direction, a neural network or score model is used to parameterize the reverse transformation. 

Diffusion models are alternatively referred to as score-based generative models since parameterizing the reverse diffusion process is equivalent to learning the `score' or $\vec{\nabla} \log p_t(x)$ of the data \citep{anderson1982reverse,song2021scorebased}. Since in high dimensions, the target distribution invariably lies on a thin manifold, the incremental addition of the random noise, blurs the distribution and makes the score progressively easier to learn. Diffusion models are the underlying mathematical framework that have given rise to the photorealistic image generation successes of \href{https://openai.com/dall-e-2}{DALL.E} and Stable Diffusion \citep{rombach2021highresolution} and have been shown to mitigate mode collapse, a phenomenon, often encountered with GANs a generative model fails to generate multiple modes in a distribution. In scientific applications, they have been used for problems involving protein folding and ligand prediction and medical imaging reconstruction \citep{corso2022diffdock,song2021solving} . They have been applied in astrophysics to reconstruction problems involving dust \citep{heurtel2023removing,mudur2022can}, cosmological simulations and initial conditions reconstruction \citep{legin2024posterior,rouhiainen2023super,ono2024debiasing,cuesta2023point,mudur2023cosmological} and strong lensing problems \citep{jagvaral2022modeling,remy2022probabilistic}.

In this work, we apply diffusion generative models to emulating cold dark matter density fields conditional on cosmological parameters and demonstrate that the trained model can also be used to derive tight and robust constraints on cosmological parameters. In Section~\ref{sec:stats}, we examine the ability of the model to appropriately capture the statistics of the distribution of fields corresponding to different parameters, and further compare the effect of modulating a single parameter at a time on the statistics of the resulting fields in the true and the generated set. We then quantify the ability of the model to capture the full range of cosmic variance for a \textit{single} parameter, as a means to assess the extent of mode collapse. In Section ~\ref{sec:pinf}, we examine how the diffusion model's approximate likelihood can be used to solve the inverse problem of constraining the cosmological parameters of a given input field. We then use the Hamiltonian Monte Carlo \citep{duane1987hybrid,neal2011mcmc,betancourt2017conceptual} method to draw samples from the estimated posterior on the cosmological parameters given an input field, and compare our estimates with a power spectrum baseline. A novel contribution of this work is our use of an HMC to sample a posterior consisting of an approximation to the diffusion model conditional likelihood to solve a downstream inference task. Finally, we demonstrate that the Diffusion-HMC based parameter inference estimates are more robust to perturbations composed of uncorrelated noise relative to the estimates from a discriminative neural network directly trained to estimate parameters.

\begin{figure*}[!ht]
  \centering
\includegraphics[width=0.95\linewidth,keepaspectratio]{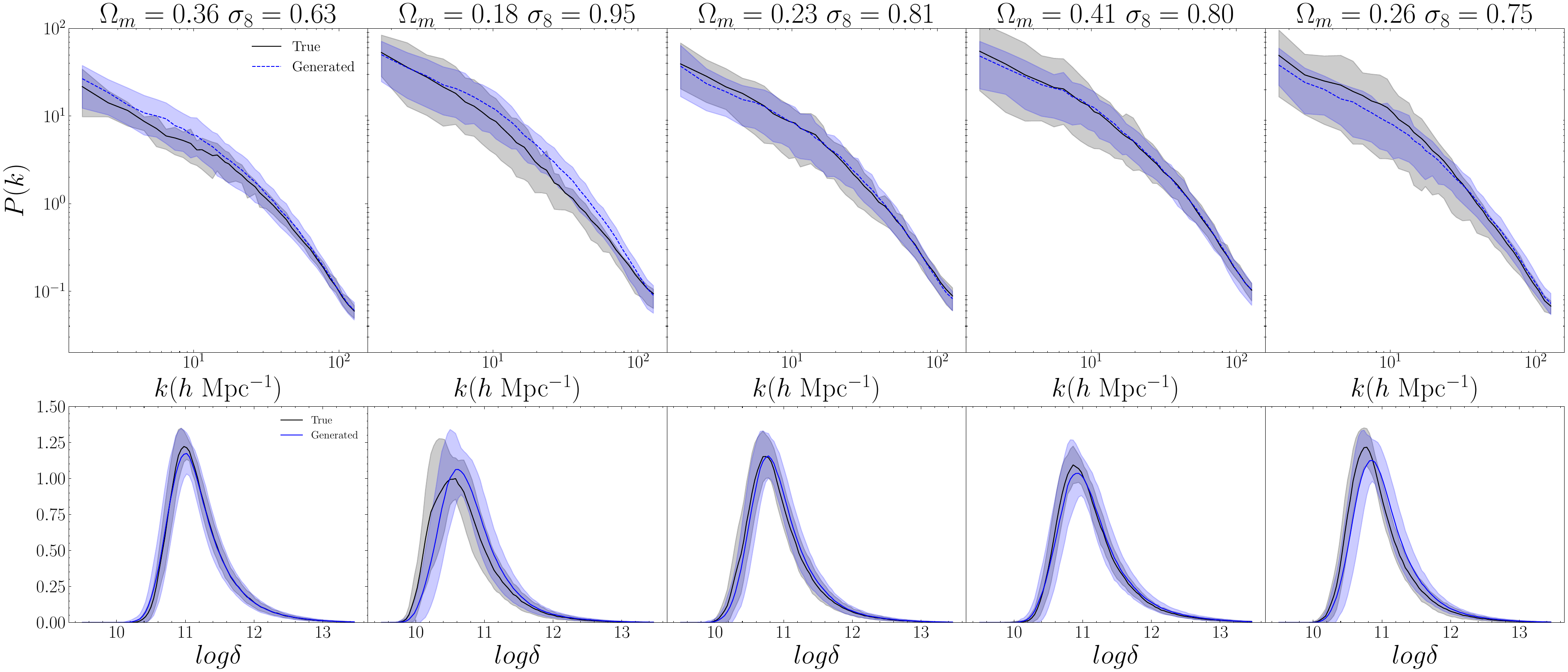}
\caption{\textbf{Generated fields at different cosmologies.} \textbf{Upper Row:} Power spectrum of the unlogged fields for five different validation parameters. The lines depict the mean power spectrum and the envelope indicates the $16^{th}$ and $84^{th}$ percentiles of the distribution. True simulations are shown in black, generated in blue. \textbf{Lower Row:} Mean and standard deviation envelopes for the density histograms of the log fields.}
\label{fig:pk}
\end{figure*}

\section{Datasets, Architecture and Training}\label{sec:training}
\begin{figure*}[!htp]
  \centering
\includegraphics[width=.95\linewidth]{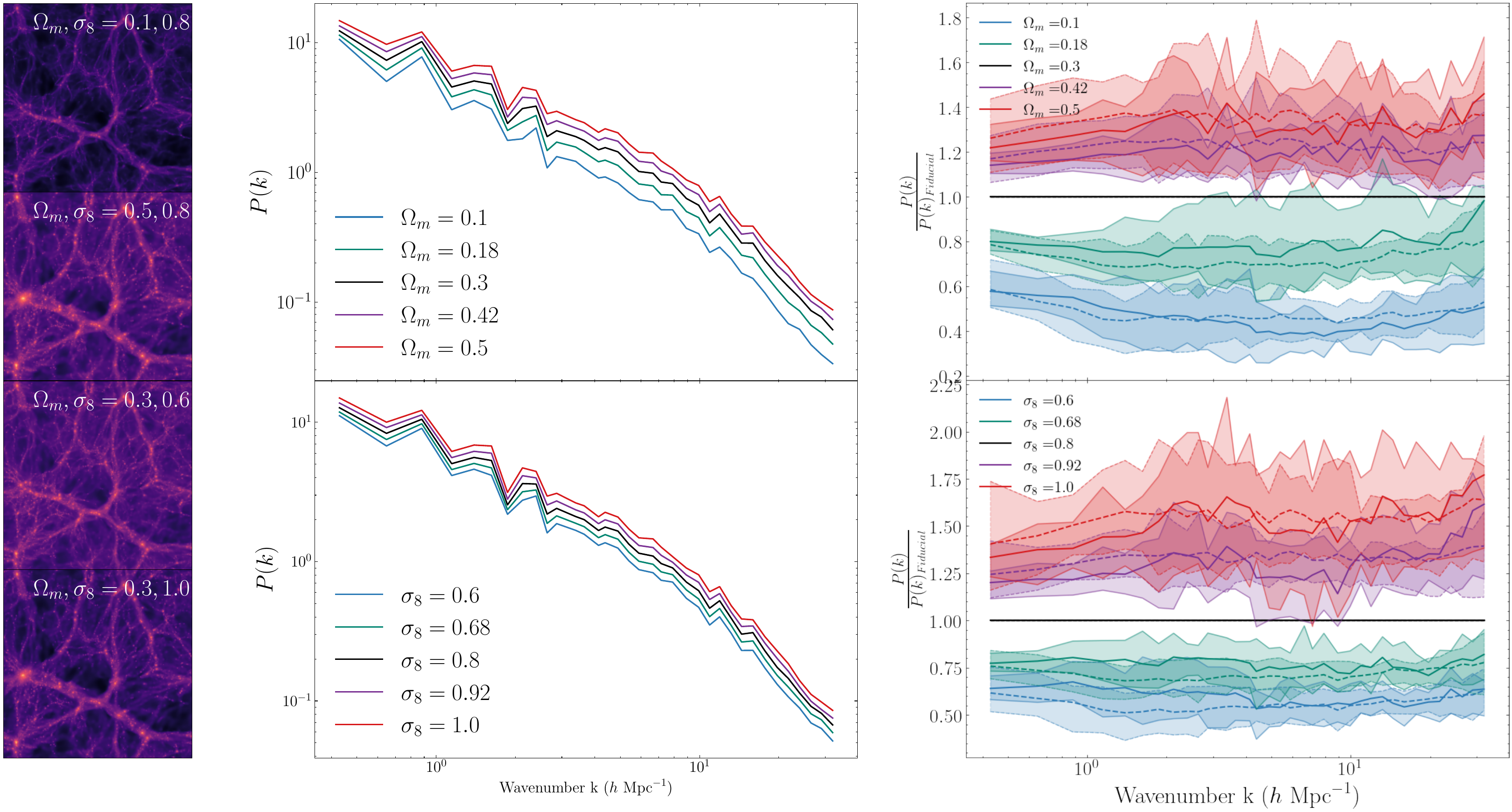}
\caption{\textbf{Generated `1P' fields.} \textbf{Left column:} Generated fields corresponding to the extreme values of each parameter for a single seed, with the other value held fixed at the fiducial value (0.3 for $\Omega_m$ and 0.8 for $\sigma_8$). \textbf{Middle column:} Power spectra of the generated fields for the same seed, for different values of each parameter, holding the other fixed. \textbf{Right column:} Mean and standard deviation for the ratio of the power spectra at the modified parameter value to the power spectra for the field at the fiducial parameter value (black) for 15 slices from the CAMELS dataset (solid) and 15 seeds for the generated fields from the diffusion model (dashed). The effect of modulating a parameter on the generated fields' power spectra is consistent with that of the true fields.}
\label{fig:conditional}
\end{figure*}

\paragraph{Datasets} We work with cold Dark Matter density fields at $z=0$ from the IllustrisTNG \citep{nelson2019illustristng, pillepich2018simulating} suite from the CAMELS Multifield Dataset (CMD) \citep{villaescusa2022camels,villaescusa2021camels}. The diffusion model is trained to generate the minmax transform applied to the log (base 10) of these dark matter fields. The minmax transform is pegged to the minimum and maximum of the log of the entire dataset, [9.42, 15.44]. The dataset contains 1000 simulations for 1000 different cosmologies with 15 two-dimensional fields per simulation. The `cosmology' is parameterized by a parameter vector with 2 cosmological ($\Omega_m$ and $\sigma_8$) and 4 astrophysical parameters. Each simulation tracked the evolution of $256^3$ dark matter particles and $256^3$ fluid elements and took around 6000 CPU hours to generate (see \cite{villaescusa2022camels,villaescusa2021camels} for more details).  The fields span $25$ \Mpc on each side. We train on 70\% of the parameters in the LH set, i.e. 700 parameters or 10,500 fields, and condition the model only on the cosmological parameters $\Omega_m$ and $\sigma_8$. Since we use dark matter density fields in this study, we do not expect them to constrain or contain much information about the astrophysical parameters. Example generated dark matter fields are shown in Figure ~\ref{fig:conditional}.

\paragraph{Diffusion Model Setup} 
\newcommand\xt[1]{\mathbf{x}_{#1}}
\newcommand\qt[1]{q(\xt{#1}|\xt{#1 -1})}
\newcommand\pt[1]{p_\theta(\xt{#1-1}|\xt{#1})}
We follow the denoising diffusion probabilistic model (DDPM) formalism, in which a target image $x_0$ is transformed to a sample from $x_T \sim \mathcal{N}(0, \mathbb{I})$ over the course of $T=1000$ timesteps. The forward diffusion process follows an incremental noise schedule \{$\beta_t$\}, and the noise is added in a variance-preserving way.
\begin{align}
& \textrm{For t} \in [0, T-1], q(x_{t+1}|x_t) = \mathcal{N}(\sqrt{1 - \beta_t}x_t, \beta_t\mathbb{I}) \nonumber \\*
& \textrm{ and } q(x_{t+1}|x_0) = \mathcal{N}(\sqrt{\bar{\alpha_t}}x_0, (1 - \bar{\alpha_t})\mathbb{I}) \nonumber \\*
& \text{ where } \bar{\alpha_t} = \prod_{t'=0}^t 1 - \beta_{t'} \label{eqn:dm}
\end{align}
The score / noise-predictor model is a U-Net \citep{ronneberger2015u} similar to that used in \cite{ho2020denoising}. There are 4 downsampling layers, with each layer consisting of 2 ResNet blocks \citep{zagoruyko2016wide}, group-normalization \citep{wu2018group}, and attention \citep{vaswani2017attention, shen2021efficient}. We use circular convolutions in the downsampling layers since the input fields have periodic boundary conditions. Each parameter is normalized to lie between [0, 1] with respect to its range, $\Omega_m \in [0.1, 0.5], \sigma_8\in [0.6, 1.0]$. A multilayer perceptron (MLP) transforms the cosmology vector into a space with the same dimension as the time embedding, and each ResNet block additionally has an MLP conditional on cosmology. The variance schedule $\beta_t$ is non-linear with smaller steps at smaller $t$ and larger steps for larger values of $t$ (see Figure ~\ref{fig:snr}). We train the model to generate the \textit{log} of the fields, and randomly rotate and flip the image to account for these invariances.
During training, for each batch of images $\xt{0}$, a batch of timesteps is sampled uniformly along with a noise pattern $\epsilon\sim\mathcal{N}(0, \mathbf{I})$. The loss function minimized is $||\mathbf{\epsilon} - \mathbf{\epsilon}_\phi(\sqrt{\bar{\alpha}_t}\xt{0} + \sqrt{1 - \bar{\alpha}_t}\mathbf{\epsilon}, t, \theta))||^2$ for each set of $\{\xt{0}, t, \theta,\epsilon\}$, where $\theta$ is the parameter vector. Our implementation minimizes the Huber loss, which behaves as an L1 loss for values of the loss greater than 1 and a Mean Squared Error (MSE) loss otherwise. From the training curves, the loss during training is less than 1 throughout, so the loss being minimized is in effect the MSE loss. We used the Weights and Biases framework \citep{wandb} to track experiments. The model has 31.2 million parameters.

\paragraph{Training} We first downsample the images by a factor of 4 and train the conditional diffusion model on these 64x64 images for 60000 iterations. Since the UNet is formulated in terms of relative downsampling, the same architecture can be applied to images with different resolution. To train the model to emulate 256x256 fields, we initialize the checkpoint with the weights of the 64x64-trained model after 60000 iterations and trained for over 400000 iterations. We found that initializing the 256x256 model with the weights of the 64x64 model and then training the model led to faster convergence.

The noise prediction loss does not fully capture sample quality and convergence, and we need an alternative metric to assess the quality of the generated samples \cite{theis2015note}. We sampled 500 fields (with 50 fields for 10 different validation parameters) for the checkpoints after 200k, 220k, 240k, 260k, 280k, 300k, 320k and 340k iterations. Sampling a batch of 50 $256\times 256$ fields from our model takes 310 seconds (6.2s/field). We computed the reduced chi-squared statistic (Equation ~\ref{eq:unichisq}) of the power spectrum of each generated field, $s$, with respect to the reference distribution comprised of the 15 true fields for that parameter. We then compute the mean and standard error of these values across all parameters and sampled fields.

While the diffusion generative model is trained to generate the fields in log space, the power spectra we compute here and in Figures ~\ref{fig:pk}-~\ref{fig:cv} correspond to the overdensity power spectra of the `\textit{linear}' ($10^{\rm{GeneratedFields}}$) fields. The checkpoint corresponding to the 260k$^{th}$ iteration had the lowest value for the chi-squared statistic of the power spectra of the linear generated fields relative to the distribution of true fields, corresponding to $2.29 \pm 0.49$ (Figure \ref{fig:ckpwise}). To put this number in perspective, we can examine the effect of cosmic variance on this metric using a leave-one-out cross-validation approach, by computing the reduced chi-squared statistic of each sample of a \textit{true} field, using the 14 other true fields corresponding to the same parameter as the reference distribution. The mean of this value across the 10 parameters is $1.70 \pm 0.36$. We use the 260k checkpoint for our analysis. We plot these values, along with the reduced chi-squared statistic of the power spectra of the \textit{log} fields and the $p$ values of the mean intensity in Appendix~\ref{app:chi2}.

\section{Summary Statistics}
\label{sec:stats}
In this section we examine the consistency of our generated fields relative to the true fields using three sets of simulations under the IllustrisTNG CAMELS suite. The Latin Hypercube, \textbf{LH}, suite varies the largest range of cosmological and astrophysical parameters but has a limited number (15) of fields for each parameter, with all parameters varying randomly. The one-parameter, \textbf{1P}, suite consists of 15 fields with the same seed, and one dimensional variations of each parameter modulated systematically, while the others are fixed to the fiducial value. The Cosmic Variance, \textbf{CV}, set consists of 405 fields for the fiducial parameter value. 

\textbf{Varying cosmology in a Latin Hypercube (LH): } We examine the consistency of the summary statistics of the distribution of true and generated fields for a given validation parameter from the LH set in Figure ~\ref{fig:pk}. We have 15 true fields and 50 generated fields for each parameter. We derived the boundaries of the envelope using the estimates of the $16^{th}$ and $84^{th}$ percentiles, while the solid line demarcates the mean of the distribution of power spectra in 35 log spaced $k$ bins. The lower panel depicts the density histograms of the log fields. The envelopes are again derived using the percentiles, while the solid lines indicate the means of the histograms for the true and generated fields. The distribution of the power spectra and the density histograms of the true and the generated fields are in good agreement with each other.

\textbf{Varying cosmological parameters one at a time (1P):} In Figure ~\ref{fig:conditional}, we generated `1P' sets and examined whether the effect of modulating a single parameter, while keeping the others constant is the same as is observed in the 1P CAMELS suite. We sample 15 fields corresponding to 15 different seeds for each of the parameters. The fields corresponding to the same seed across parameters have the same position and orientation of their seeded structures as visible in Figure ~\ref{fig:conditional}. For each seed, we then compute the ratio of the power spectrum of a field at a different parameter to the power spectrum of the fiducial parameter value, and compute the average and the percentile-based standard deviation envelopes across all seeds as depicted in the right column of Figure ~\ref{fig:conditional}. The dashed (solid) line and envelope correspond to the generated (true) ratios. The ratios for the generated `1P' set are in good agreement with those of the true `1P' set, since the dashed lines are within the solid envelopes.

\textbf{Reproducing cosmic variance (CV):} The CV set has 405 ($27\times15$) fields for the fiducial parameter value of [0.3, 0.8] and varying initial conditions, designed to quantify the effect of cosmic variance. The CV set allows us to quantify the consistency between the second moments of the true and the generated distributions. In particular, it allows us to test the ability of the model to generate a diverse set of samples for the same cosmological parameters such that it reproduces the true underlying distribution at fixed cosmology. 

\begin{figure*}[htp]
  \centering
  \includegraphics[height=0.24\textheight]{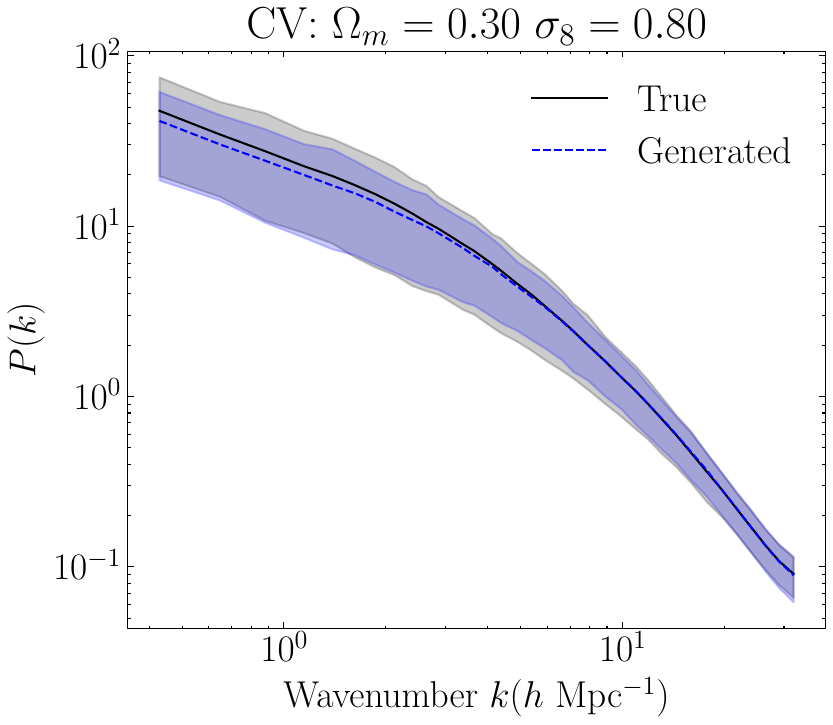}
  \includegraphics[height=0.24\textheight]{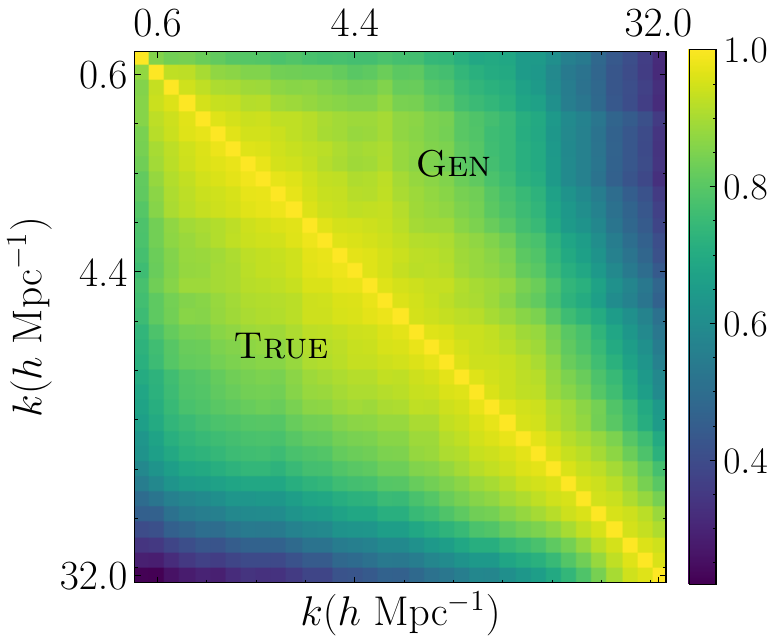}
  \begin{minipage}[b]{0.32\linewidth}
\end{minipage}
\caption{\textbf{Generated `CV' fields.} \textbf{Left:} Power spectra of the 405 true and generated fields, with the mean, and $16-84$th percentiles. \textbf{Right:} Correlation matrix of the power spectra of the true and generated fields. The lower triangular matrix corresponds to the correlation matrix of the true fields while the upper triangular matrix corresponds to the correlation matrix of the generated fields.}
\label{fig:cv}
\end{figure*}

\begin{figure}[htp]
  \centering\includegraphics[width=0.75\linewidth,keepaspectratio]{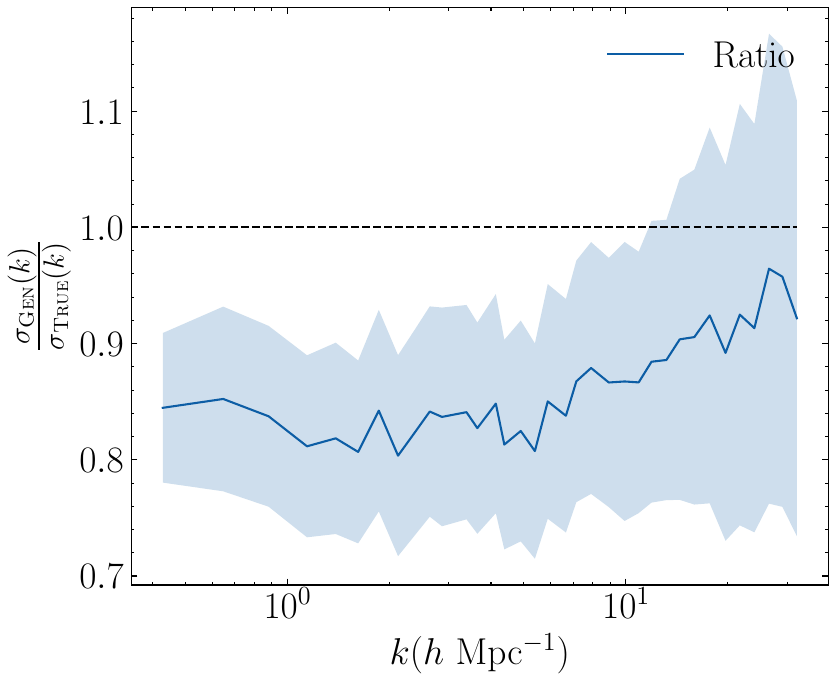}
\includegraphics[width=0.75\linewidth,keepaspectratio]{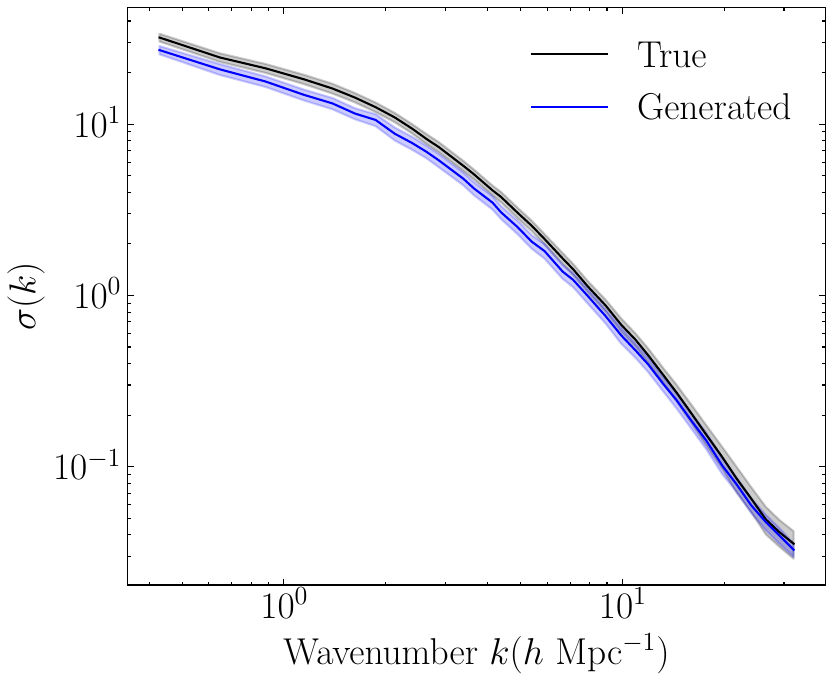}
\caption{\textbf{Generated `CV' fields.} \textbf{Upper:} Ratio of the standard deviations of the generated fields to that of the true fields. \textbf{Lower:} Standard deviations of the power spectra in each $k$ bin for the generated and true fields.}
\label{fig:cv_std}
\end{figure}

We generate 405 samples from our trained diffusion model, compute their power spectra, and examine the standard deviation, and the correlation matrices of different $k$ modes of the power spectrum in Figures ~\ref{fig:cv} and ~\ref{fig:cv_std}. The correlation between the modes of the power spectra is largely consistent between the true and the generated samples, although the generated samples appear to have a slight excess correlation around 5 \hbyMpc. 

The standard deviations of the distribution are also consistent although the standard deviations of the generated power spectra are slightly underpredicted relative to the true power spectra at the largest length scales ($k<5$\hbyMpc). To construct the standard deviation estimator, we use the jackknife-approach to generate a distribution of 405 estimates of the standard deviation for each set, where the $\rm{i^{th}}$ estimate corresponds to the standard deviation computed using all samples excluding that of the $\rm{i^{th}}$ field. We then compute the mean and the standard deviation for these sets in order to capture the mean and the standard deviation on the estimate of the standard deviation in the upper right panel of Figure ~\ref{fig:cv}. We use the ratios of the jackknife-estimated means and errors in the lower right panel.

The ability to capture the full diversity of $k$ modes for a single parameter may be in tension with the ability to distinguish between and appropriately modulate the power spectrum for different parameters. \change{This tension is enhanced for our assessment of mode collapse in spectral space, compared to canonical machine learning datasets involving discrete classes, since the cosmological parameters that we condition the diffusion generative model on lie on a continuum. Thus, for a given conditioning parameter, a generated field with too much or too little power on certain scales relative to the mean of the statistic is more likely to wander into the typical set of a field with a different cosmological parameter since modulating the parameters also modulates the power spectrum (as in Figure ~\ref{fig:conditional}). In the context of natural images, one often deals with categorical descriptors, where the boundaries between whether an object qualifies as one class or another are typically more clearly delineated, and the ability to capture the full diversity of cats is unlikely to cause the model to `wander' into islands of image space corresponding to a dog or an airplane. Thus the slightly lower standard deviation can be partly attributed to the possibility that the model chooses to compromise on the diversity of samples for a single parameter, in order to be able to accurately generate fields that look different for different parameters.}


We compute the covariance between the modes of the power spectrum. With the full covariance we can now statistically quantify the consistency of the generated samples relative to that of the true samples using the multidimensional reduced chi-squared statistic. We use 350 samples of the true fields to set up the reference distribution used to compute the covariance. We compute the inverse covariance and adjust for the Hartlap factor (Equation ~\ref{eq:hartlap}) \citep{hartlap2007your}. We then compute the multidimensional reduced chi-squared statistic of the entire sample of the true and generated fields, and compute the means of the 405 chi-squared statistics for each distribution following Equation ~\ref{eq:multidim_chisq}. For the true fields, the mean of the chi-squared distribution is 33.4 while that of the generated is 36.6. Since our power spectra consist of 35 log-spaced bins, this is consistent with the expected mean for a chi-squared distribution with 35 degrees of freedom, i.e. 35.


\newpage
\section{Parameter Inference}
\label{sec:pinf}
A trained diffusion model can be used to estimate the variational lower bound (VLB) of the log likelihood \citep{kingma2023understanding, cuesta2023point}. In the case of a conditional diffusion model, this likelihood estimate is also conditional, i.e. 

\newcommand{\lzero}[1]{{\color{teal}#1}}
\newcommand{\lt}[1]{{\color{red}#1}}
\newcommand{\lT}[1]{{\color{blue}#1}}

\begin{align}
& \mathbb{E}_q[-\log p_\phi(x_0|\theta)] \leq \mathbb{E}_q [\lzero{-\log p_\phi (x_0|x_1, \theta)} + \nonumber \\*
& \lt{\sum_{t\geq1} \rm{D_{KL}}[q(x_{t}|x_{t+1}, x_0) || p_\phi (x_t|x_{t+1}, \theta)]} + \nonumber \\*
& \lT{\rm{ D_{KL}} [q(x_T|x_0) || p(x_T)]}] \simeq L_{\rm{VLB}} \label{eqn:exact_vlb} \\*
& L_{\rm{VLB}} = \lzero{L_0} + \lt{L_1 ... L_{T-1}} + \lT{L_T} \\*
& \hat{L}_{\rm{VLB}} = \sum_{t< T_{\rm{MAX}}} L_t \label{eqn:approx_vlb} 
\end{align}

where $\phi$ denotes the diffusion model architecture, $\theta$ is the conditioning cosmology (in our case, a vector with $\Omega_m$ and $\sigma_8$), $p_\phi$ are the reverse (learned) distributions, and $q$ are the forward (analytical) distributions. During training, the diffusion model's noise prediction loss terms are equivalent to terms of the reweighted VLB \citep{kingma2023understanding}. Since the predicted noise is conditional on cosmology, the terms of the variational lower bound thus \change{encode dependencies on the cosmological parameters}. The contrast between the VLB evaluated at one parameter $\theta_1$ relative to another parameter $\theta_2$ for a \textit{fixed} field $x_0$ can thus be used to find the region of parameter space that maximizes the conditional likelihood for the field. 

\subsection{Examining the influence of different timesteps}
\begin{figure*}[!ht]
\centering
\begin{minipage}[b]{0.4\linewidth}
  \centering
  \includegraphics[width=\linewidth]{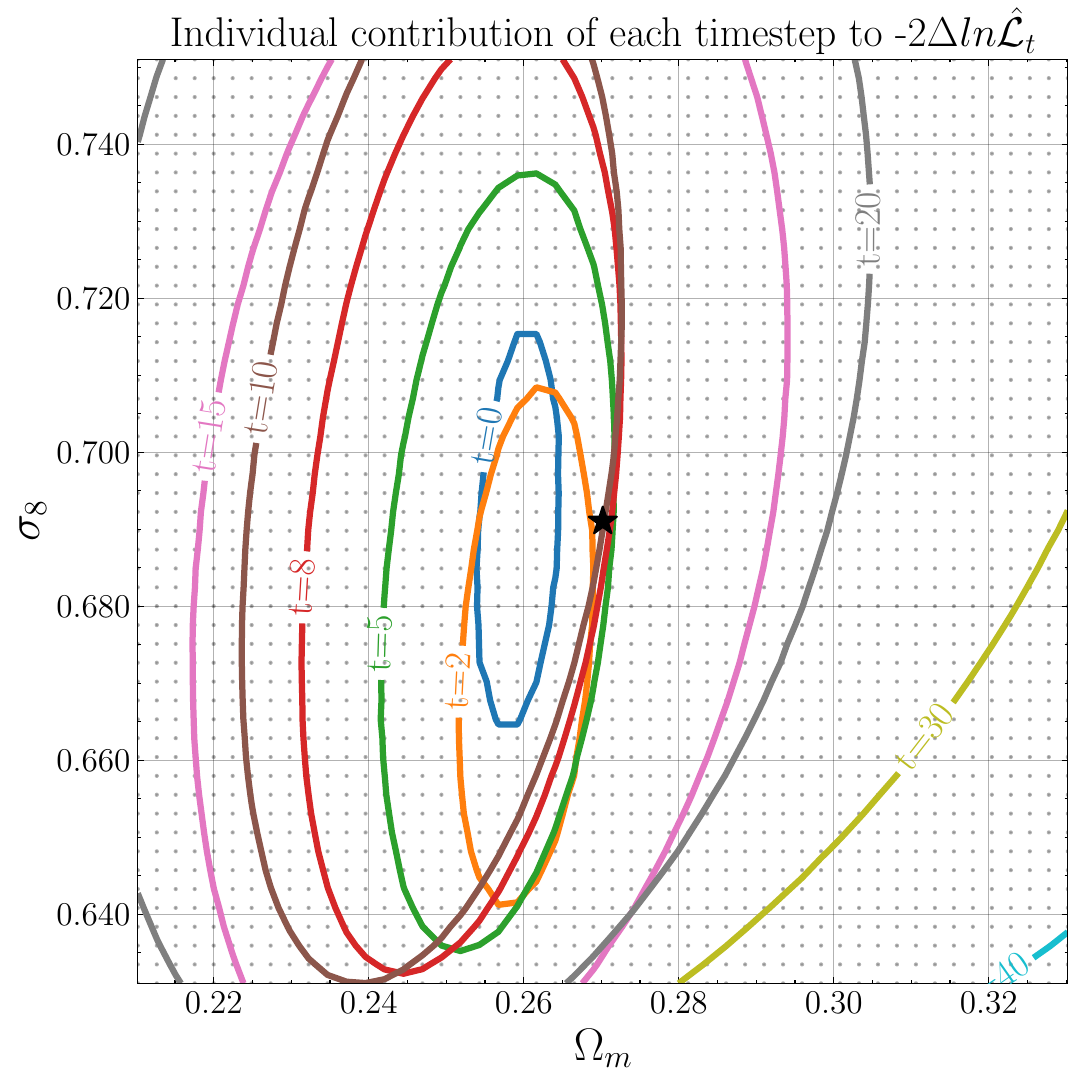}
\end{minipage}%
\hfill
\begin{minipage}[b]{0.57\linewidth}
  \centering
  \includegraphics[width=\linewidth,height=100pt]{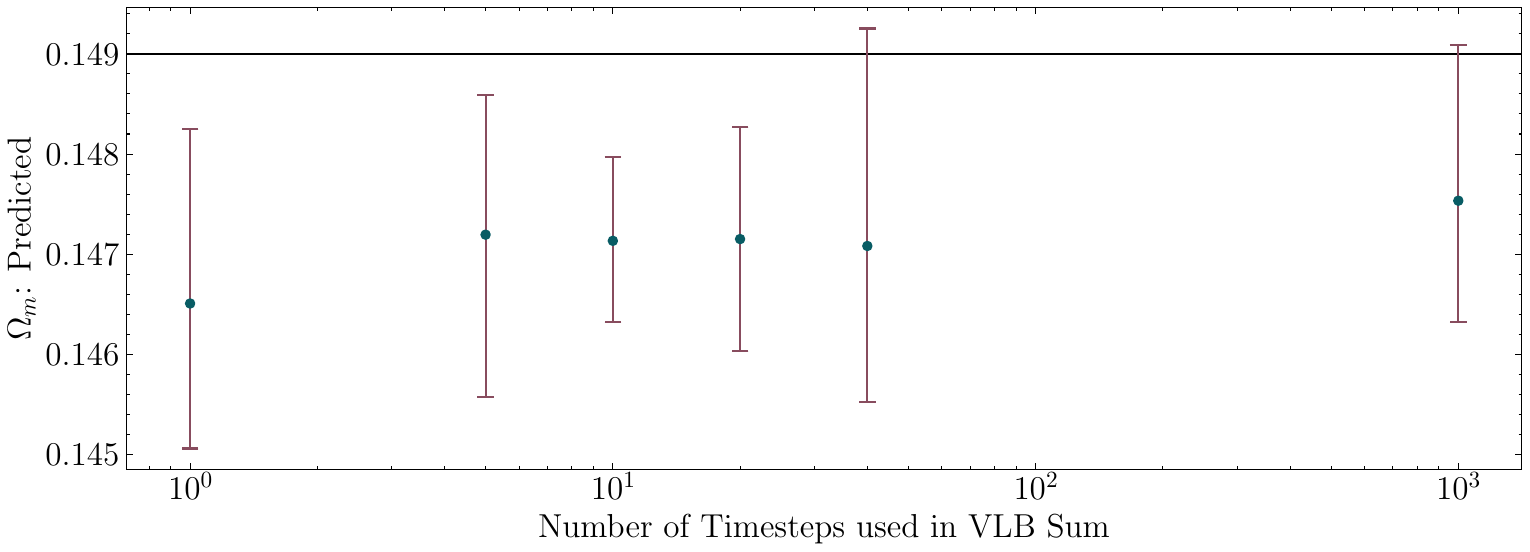}\\[\abovecaptionskip]
  \includegraphics[width=\linewidth,height=100pt]{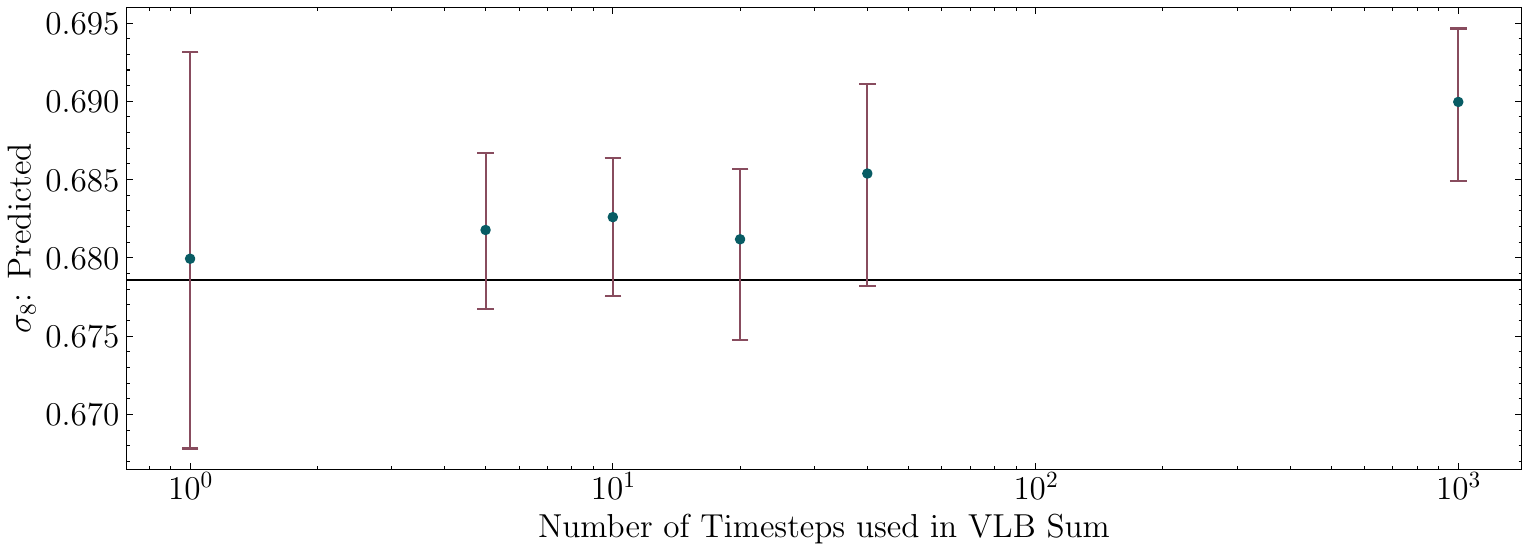}
\end{minipage}
\caption{Investigating the contribution of the timesteps used in the VLB sum. \textbf{Left:} One sigma contours of each $-2\Delta ln L_t$'s \textit{individual} contribution for different timesteps. The contours are all centered near the true parameter (black star) and become wider as t increases. \textbf{Right:} Mean, and 1 sigma predictions for a single field as a function of the number of timesteps used in the VLB sum \change{optimized by} the HMC. Reducing the number of timesteps used to compute the VLB does not significantly affect the diffusion model predictions.}
\label{fig:clik_t}
\end{figure*}

Since using all the terms $L_t$ that contribute to the VLB is computationally expensive, we first investigate how sensitive each of the contributing terms $L_t$ is to changes in cosmological parameters over a grid in Figure ~\ref{fig:clik_t}a). For an input field $x_0$ and a single seed, we evaluate each of the $L_{t}(x_0|\vec{\theta}_{\textsc{Eval}})$ terms over a $50\times50$ grid in $[\Omega_m, \sigma_8]$, centered on the value of the true field $\vec{\theta}_{\textsc{True}}$ and extending to $\pm 0.06$ about the true parameter. To disentangle the \textit{individual} contribution of each term, we subtract the minimum value of each $L_t(x_0|\vec{\theta}_{\textsc{Eval}})$ on the grid, and multiply it by 2 to yield $-2\Delta ln\mathcal{\hat{L}}_t$. We plot the contour corresponding to 2.30, or the one-sigma contour for a chi-squared distribution with two degrees of freedom. Two observations are apparent: all contours are minimized in the vicinity of the true parameter, and the curvature of the contours decreases with increasing time. Increasing time involves increasing the amount of noise added to the input image, which can explain the increased uncertainty in the true value of the parameter. Thus dropping the terms corresponding to the higher timesteps in Equation~\ref{eqn:exact_vlb} is unlikely to result in weaker constraints on cosmology. 

\subsection{HMC-based Parameter Inference}
\newcommand{\mom}{\bm{p}}
To compute the parameter estimates for fields, we draw samples from the posterior on the parameter using the Hamiltonian Monte Carlo (HMC) method. \change{The Hamiltonian Monte Carlo, or Hybrid Monte Carlo method \citep{duane1987hybrid,neal2011mcmc,betancourt2017conceptual} is a Markov Chain Monte Carlo (MCMC) approach that draws samples from a probability distribution $\pi(\theta)$ via the introduction of an auxiliary momentum variable $\mathbf{p}$ and solves the equations of Hamiltonian dynamics in order to update the momentum and position ($\theta$). HMC enables a more efficient exploration of high dimensional probability distributions.} \change{In our case, using an HMC also helps us circumvent the problem of having to redefine the extents and granularity of the parameter grid depending on how confident the constraints for a given $T_{\rm{MAX}}$ are.} The Hamiltonian governing the dynamics in the HMC chain is: 

\begin{align}
\label{eq:hmc}
& H = U + K \text{ where } U = - \log \pi(\theta) \text{ and }  K = \frac{\mom^{-1} \bm{M}^{-1}\mom}{2} \\*
& \log \pi(\theta) = \log p_\phi(x_0|\theta) + \log p_{\textsc{Prior}} (\theta)  \nonumber \\*
& \simeq -\hat{L}_{\rm{VLB}} + \log p_{\textsc{Prior}} = -\sum_{t< T_{\rm{MAX}}} L_t + \log p_{\textsc{Prior}}
\end{align}
We used the \texttt{Hamiltorch} \citep{cobb2019introducing} package for the HMC. We \change{explain more details} about the HMC implementation in Appendix ~\ref{app:hmc}. The prior is chosen to be a flat prior over $\Omega_m \in [0.1, 0.5]$ and $\sigma_8 \in [0.6, 1.0]$. The initial parameter value is always the fiducial value of $[0.3, 0.8]$ and we designate the first 100 samples as burn-in samples that are discarded. We reduce the averages over random noise in each $L_t$ term in Equation \ref{eqn:exact_vlb} to a single stochastic estimate of $L_t$ in Equation \ref{eqn:approx_vlb}.

We now examine the effect of truncating terms in Equation ~\ref{eqn:approx_vlb} using the HMC based parameter inference. \change{Truncating terms allows us to perform inference faster and can allow us to explore the tradeoff between dropping terms for speed and higher precision with more timesteps.} In the right column of Figure ~\ref{fig:clik_t}, for a single field, we plot the mean predictions and the $15.9-84.1$ (1 sigma) percentiles computed using 200 samples with the approximate $-\log p_\phi(x_0|\theta)$ using Equation~\ref{eqn:approx_vlb} as a function of $T_{\rm{MAX}}$ on the x axis for the same field. The true value of the cosmological parameter for this field is demarcated by the solid black lines. For this field and parameter, using more terms asymptotically removes the bias on $\Omega_m$ while increasing the bias on $\sigma_8$. However, using the first 20 timesteps only changes the mean prediction for $\Omega_m$ and $\sigma_8$ by $-0.26\%$ and $-1.27\%$ of the prediction using all 1000 timesteps respectively. 

\begin{figure*}
\centering\includegraphics[width=0.55\linewidth]{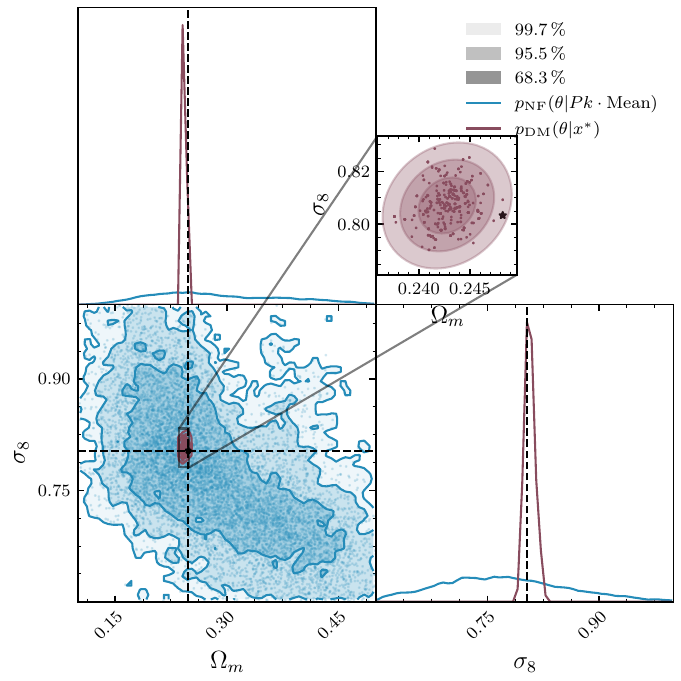}
\includegraphics[width=0.45\linewidth]{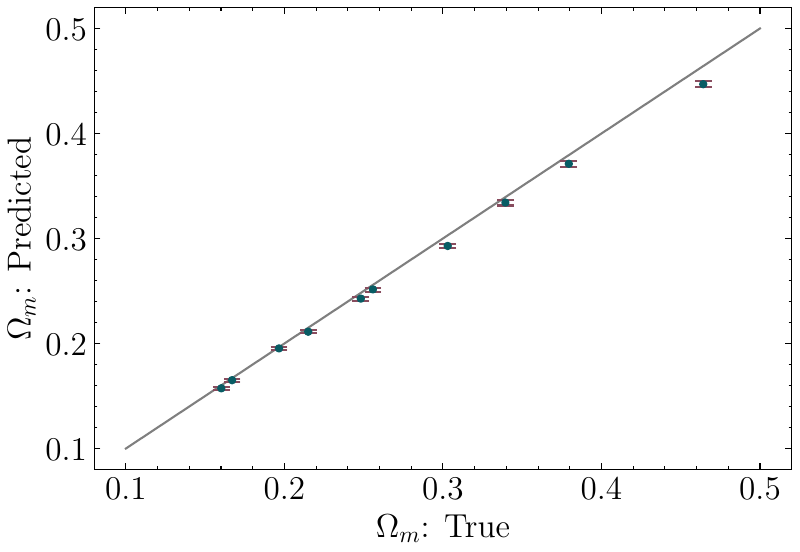}
\includegraphics[width=0.45\linewidth]{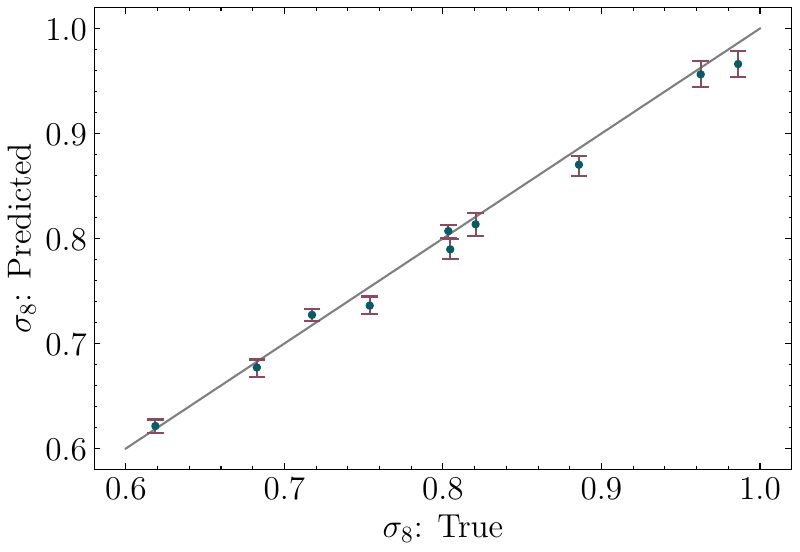}
\caption{\textbf{Top:} Comparison of the parameter estimates for a single field with the power spectrum based estimator and with the diffusion model likelihood. The star demarcates the true parameter corresponding to the input fields. The shaded contours demarcate the $68.27\%, 95.45\%$ and $99.73\%$ confidence intervals. 
\textbf{Lower Panel:} Predicted parameter and ground truth parameter for 10 different fields. The error bars correspond to the $15.9-84.1$ percentiles for the marginal probability distributions for each parameter for each input field.} \label{fig:hmc_main}
\end{figure*}

We now turn our attention to the performance of our parameter inference approach relative to a power spectrum baseline. For subsequent HMC-based parameter inference in this section, we use Equation~\ref{eqn:approx_vlb} with $T_{\rm{MAX}} = 20$ to approximate the conditional negative log likelihood. Drawing 500 samples with $T_{\rm{MAX}}=20$ takes $\sim32$ minutes for a single field. 

To assess the constraining power of the power spectrum, we use the neural posterior estimation approach and train a normalizing flow to represent the posterior of $p_{\rm{NF}}(\theta|Pk\cdot\rm{Mean})$ using the \texttt{Lampe} \citep{rozet2021lampe} package. Since computing the overdensity power spectrum involves dividing by the mean of the field, we further concatenate the mean of the log field as an additional feature, since the prediction for $\Omega_m$ for a small box size of $25$ \Mpc \change{is very sensitive to the mean of the fields}. For a single field, we compare the posteriors obtained by drawing 10000 samples from the power spectrum based estimator and 400 samples from the diffusion model+HMC in Figure ~\ref{fig:hmc_main}. Other details about our implementation are in the Appendix ~\ref{app:bl}. 

The diffusion model has significantly narrower constraints for the parameters relative to the power spectrum baseline. \change{Note that, as shown in Figure~\ref{fig:conditional}, the cosmological parameters $\Omega_m$ and $\sigma_8$ are strongly correlated at the level of the power spectrum on the small scales probed by our simulations, since they are both modulating the amplitude of the power spectrum.} This result is consistent with \cite{villanueva2022learning} who also found that the galaxy power spectrum in conjunction with a multilayer perceptron yielded weak constraints on $\Omega_m$ and $\sigma_8$.

In the second row, we plot the predicted cosmological parameters relative to the truth for 10 different fields across different parameters in the validation set. The dots annotate the mean of the 400 samples and the error bars correspond to the $15.9-84.1$ percentiles of the samples. The mean of the samples is close to the true value of $\vec{\theta}_{\textsc{True}}$ over a broad range of parameters. The $\Omega_m$ predictions are biased by $-0.006$ $(2\%$ of the true prediction) on average over the range of parameters while the mean absolute bias for $\sigma_8$ is 0.01 ($1.26\%$ of the true prediction). The $\sigma_8$ prediction uncertainties (mean z score $\bar{|z|} = 1.14$) are better calibrated relative to the uncertainties for $\Omega_m$ ($\bar{|z|} = 2.75$), \change{but overall, we find the error bars to be slightly under-predicted. We attribute the mis-calibration to the truncation of terms. It is possible that including more timesteps, averaging over more seeds and examining alternate choices of the variance schedule could yield better calibrated uncertainties; we leave this exploration for future work.}

\begin{figure*}[!htp]
\centering
\includegraphics[width=0.8\linewidth]{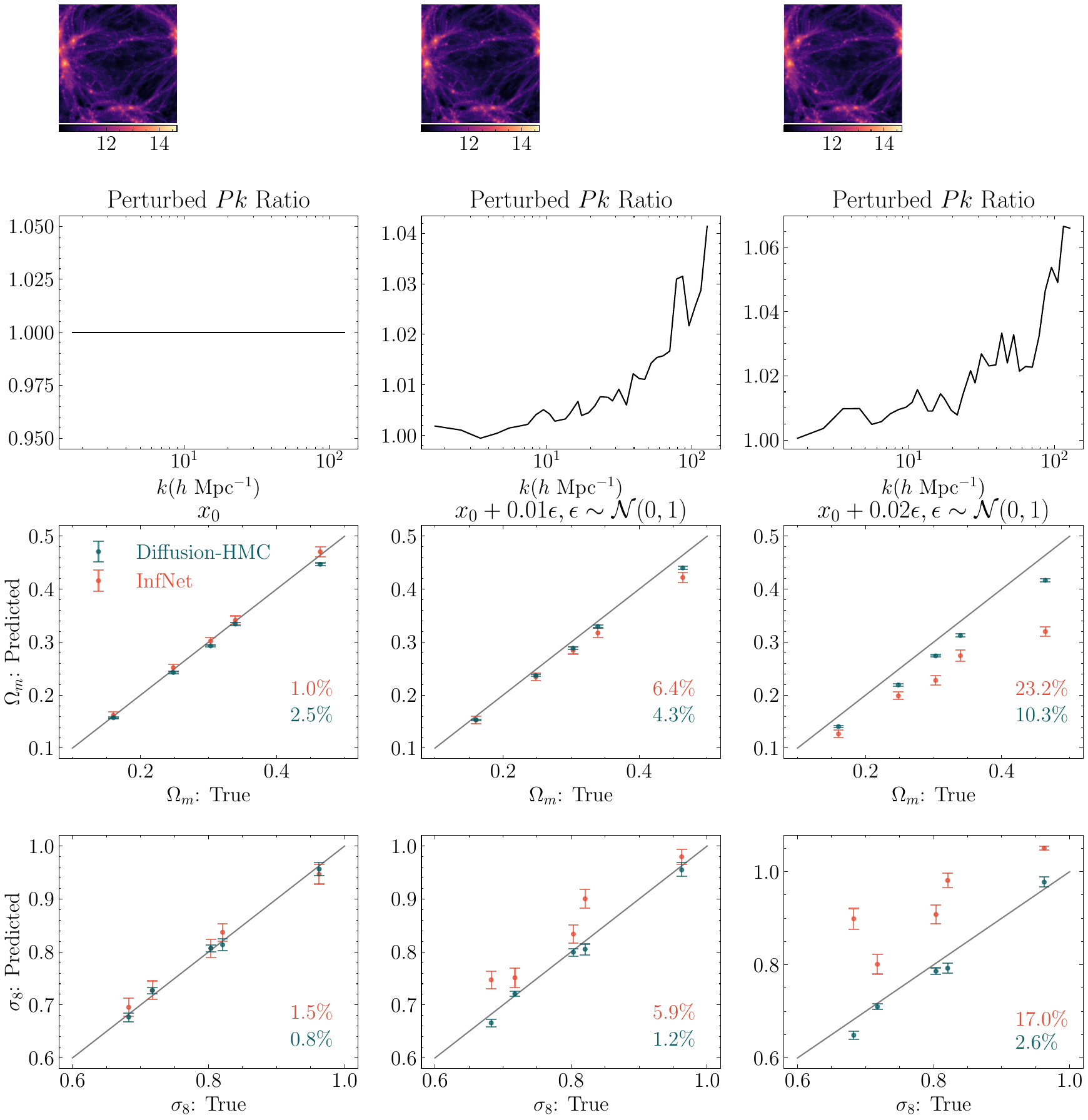}
\caption{\textbf{First row:} An example field corresponding to $\Omega_m, \sigma_8 = 0.33, 0.96$ with its perturbed counterparts, along with the ratio of the perturbed \textbf{linear} field's power spectrum to that of the original field $\frac{Pk_{Noisy}}{Pk_{Original}}$. \textbf{Second, Third rows:} Mean and $15.8-84.1$ percentile predictions using the Diffusion-HMC estimates (green) and the mean and sigma predictions obtained from the parameter inference network in \cite{villaescusa2021multifield}, denoted by InfNet. The mean absolute percentage bias (of the true prediction) for each parameter is also indicated. The Diffusion-HMC constraints are more robust to perturbations to the input image.}
\label{fig:robust}
\end{figure*}

\subsection{Robustness}
\change{Although our constraints are much tighter than those derived from the two point correlation function, robustness to noise and observational systematics is an important consideration guiding the use of parameter inference methods on survey data.} Since the terms of the VLB (and the noise prediction loss terms) include terms where the original image has been noised, we hypothesize that the parameter estimates learned by the model are \textit{naturally} more robust to perturbations involving the addition of scaled white noise to the field. 

To test this, we examine the extent to which our parameter estimates change relative to the predictions of the parameter inference network in \cite{villaescusa2021multifield}. The neural network in \cite{villaescusa2021multifield} is trained to predict the mean and the standard deviation of the parameters given an input dark matter field. In the leftmost column of Figure ~\ref{fig:robust}, we compare sample predictions obtained from the diffusion model ({\color{teal}Diffusion-HMC}) relative to those obtained from the parameter inference network ({\color{red}InfNet}) on 5 fields ($x_0$). In the next two columns, we add increasing levels of $\mathcal{N}(0, \sigma)$ noise with $\sigma \in [0.01, 0.02]$, and examine the estimates for both approaches. We perform this experiment in a noise-agnostic setting, i.e. we assume that we do not know the level of noise a priori and do not modify either the Diffusion-HMC approach or the parameter inference network based approach. We use the same $T_{\rm{MAX}}=20$ and use 200 samples for the Diffusion-HMC estimates for all fields. Although the noise perturbations to the field are visually imperceptible, and correspond to a change of less than $\sim 4-6\%$ to the power spectrum at the smallest scales, the predictions of the parameter inference network are significantly ($\sim 6-23\%$) disturbed. Indeed, we find that the Diffusion-HMC based constraints are perturbed far less than the InfNet constraints. The Diffusion-HMC $\sigma_8$ constraints are also more robust than its $\Omega_m$ constraints. 

\section{Conclusion}
In this work, we used a diffusion generative model to emulate dark matter density fields conditional on cosmological parameters. The model learns to reproduce the modulation of summary statistics, such as the power spectra, for changes in cosmological parameters. We additionally assess the extent of mode collapse through the lens of cosmic variance by examining the diversity of the power spectra at the fiducial cosmology. 

We then directed our attention to the inverse problem of parameter inference and disentangled the contribution of each term $L_t$ in the expression for the Variational Lower Bound to constraints on cosmological parameters. Our findings reveal that the strength of the constraints decrease as $t$ increases. \cite{clark2024text} and \cite{li2023your} explored this question in the context of using the diffusion model's conditional VLB terms for classification. \cite{clark2024text} also found a negative exponential weighting of the terms to be optimum while \cite{li2023your} found that intermediate timesteps had the highest accuracy when only a single timestep is used. The difference in which terms contain the most information about the conditioning attribute (cosmology / class) is interesting, and could be partly attributed to the difference in their formulation and weighting of the VLB terms and in how changes in the conditioning vector affect an image in different settings. \change{Modulating cosmology modifies global attributes of the fields, such as their intensity and power spectra, as seen in Figure ~\ref{fig:conditional}, while the information that distinguishes different breeds of pets from each other tends to be relatively localized in the bounding box containing the animal}. 

If it is indeed always the case that the timesteps nearest to the image manifold contain most cosmological information, one could in future, swap out our discrete time architecture with continuous time diffusion models \citep{song2021scorebased}, where $t$ is a continuous variable with $t\in [0, 1]$ and prioritize steps that lie near the image manifold or $t=0$. 

These insights motivated us to truncate the diffusion model conditional VLB based approximation for $p_\phi(x_0|\theta)$ by subsampling the terms to only use the first $T_{\rm{MAX}}$ terms (Equation ~\ref{eqn:approx_vlb}). This approximation allows us to backpropagate its gradient and plug this estimate for $p_\phi(x_0|\theta)$ into an HMC and sample the posterior on $p_\phi(\theta|x_0)$. The Diffusion-HMC approach yields tight constraints on cosmological parameters, competitive with a bespoke parameter inference network \cite{villaescusa2021multifield} trained only to infer cosmology given a field. 

In our experiments, we use only a single seed in each HMC step and a $T_{\rm{MAX}}=20$ to speed up inference. \change{In our case, the use of an HMC allowed us to circumvent the requirement of choosing a grid with the appropriate resolution required to resolve the constraints, and eliminates the dependence on the number of points in the grid. However, the Diffusion-VLB estimates could also be used in other settings, that do not entail the use of an HMC.} While our approach scaled with the number of timesteps we use in the sum, in constrast, \cite{li2023your} scaled with the number of classes in the classification dataset. 

Lastly, we demonstrate that the diffusion model likelihood confers the Diffusion-HMC constraints with greater robustness against the addition of small amounts of noise to the input image relative to the behavior of a parameter inference network. This echoes the behavior of diffusion models in other discriminative tasks such as \citep{li2023your,chen2024your,prabhudesai2023diffusion}, where diffusion model based classifiers have been shown to possess higher robustness to adversarial examples or perturbations. \change{This is a pertinent finding since \cite{horowitz2022plausible} showed that the powerful constraints derived from neural network based parameter inference may often come at the cost of their susceptibility to slight perturbations that the canonical two-point correlation based analyses are impervious to.} We find that our diffusion-model based parameter inference approach enables more noise-robust field-level inference. It would be interesting to further explore the differences in inductive biases learned by discriminative networks (e.g.: \citep{villaescusa2021multifield, sharma2024comparative}) relative to generative models repurposed for discriminative tasks.

The simulation volume of the dataset we work with is still much smaller than the scales mapped by astrophysical surveys. Future work could focus on scaling up to more survey-realistic scenarios involving larger simulation volumes and directly observed tracers. \cite{cuesta2023point} also showed that diffusion generative models that work with point clouds can allow one to emulate and perform cosmological parameter inference with point cloud data. Our exploration into robustness \change{could inform} applications to real data, and future investigation could focus on conferring and quantifying robustness against other survey-related and observational noise effects. Alternative formulations of the generative process \citep{bansal2024cold,wildberger2024flow} could also be relevant to this exercise. 

In this work, we demonstrated that a diffusion model can be trained not just to emulate fields, but that it's likelihood estimate can be adapted to work with the Hamiltonian Monte Carlo framework to derive tight constraints on cosmological parameters. This makes a step toward advancing the use of diffusion model based priors for a range of downstream tasks from image generation and restoration to inference problems.

\section{Code}
The checkpoint used for the results in this paper is available on Zenodo: doi:
\href{https://doi.org/10.5281/zenodo.13993010}{10.5281/zenodo.13993010}. The code for this project is available at \href{https://github.com/nmudur/diffusion-hmc}{Diffusion-HMC} \github{ with} a current copy uploaded to the above Zenodo DOI. We acknowledge \href{https://huggingface.co/blog/annotated-diffusion}{The Annotated Diffusion}, \href{https://github.com/openai/improved-diffusion}{Improved-Diffusion} \citep{nichol2021improved}, \href{https://github.com/google-research/vdm}{VDM} \citep{kingma2021variational}, \href{https://github.com/hojonathanho/diffusion}{DDPM} and \cite{song2021scorebased} for code snippets and blocks used in the diffusion model architecture and \cite{villaescusa2021multifield} for the parameter inference baseline code snippets and networks.

\textit{Packages:} \texttt{Hamiltorch} \citep{cobb2019introducing}, \texttt{Lampe} \citep{rozet2021lampe}, GNU Parallel \citep{tange_ole_2018_1146014}.

\change{
\section{Acknowledgements}
This work was supported by the National Science Foundation under Cooperative Agreement PHY2019786 (The NSF AI Institute for Artificial Intelligence and Fundamental Interactions). We are grateful to Yueying Ni, Francisco Villaescusa-Navarro, Core Francisco Park, Andrew K. Saydjari, Justina R. Yang, Yilun Du, Ana Sofia Uzsoy, Shuchin Aeron and many others for insightful conversations.}

\bibliography{sample}{}
\bibliographystyle{aasjournal}

\clearpage
\onecolumngrid
\appendix
\section{Parameter Inference}
\begin{figure*}[h!]
\centering
\includegraphics[width=0.32\linewidth]{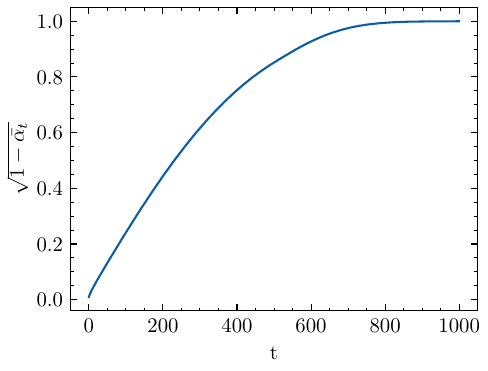}
\includegraphics[width=0.32\linewidth]{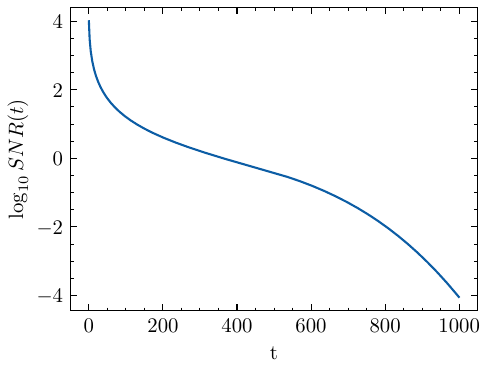}
\caption{The scale / standard deviation of the cumulative noise in Equation ~\ref{eqn:dm} added to the image over different timesteps.}
\label{fig:snr}
\end{figure*}

\label{app:hmc}
In Equation ~\ref{eq:hmc}, we set $\bm{M}^{-1}$ to be diagonal with [1, 5]. Setting the inverse mass matrix in an HMC to be close to the covariance of the expected posterior distribution helps the chain explore the space better. In this case, we choose a step size of $5\times10^{-4}$, because of the steep gradient of the posterior distribution with respect to $\Omega_m$. For the parameter on field where the true parameter is 0.101, i.e. on the prior range, we need to further reduce the step size to $1\times10^{-4}$ in order for the parameters to be accepted. We modified the \texttt{Hamiltorch} package in order to generate samples. For $T_{\rm{MAX}}\geq20$ we compute the contributions to the VLB loss in batches of 10 timesteps and accumulate the gradient contribution for each batch. This enables us to compute the gradients with 1000 timesteps. The choice of mass matrix accelerates the chain's convergence to the correct region of parameter space for $\sigma_8$. We approximate -$\log p_\phi(x_0|\theta)$ with the truncated variational lower bound in Equation ~\ref{eqn:approx_vlb}. While we do not compute the expectation over multiple seeds \textit{within} a single evaluation of Equation ~\ref{eq:hmc}, for speed, every evaluation uses a different seed and noise pattern. The prior is chosen to be a flat prior over $\Omega_m \in [0.1, 0.5]$ and $\sigma_8 \in [0.6, 1.0]$. 

\begin{align}
& L_{vlb} = L_0 + L_1 ... L_{T-1} + L_T = \\* \nonumber
& = \mathbb{E}[-\log p_{\phi} (x_0|\theta)] \leq \mathbb{E}_q [\rm{ D_{KL}} [q(x_T|x_0) || p(x_T)] + \sum_{t\geq1} D_{KL}[q(x_{t}|x_{t+1}, x_0) || p_\phi (x_t|x_{t+1}, \theta)] - \log p_{\phi} (x_0|x_1, \theta)]
\end{align}

\newcommand{\pvec}{\vec{\phi}}
\newcommand{\cdm}{\textrm{DM}_{\theta|\pvec}}
\textsc{Notation:}
\begin{itemize}\itemsep0em
    \item $x_0$: Normalized input field
    \item $\phi$: Noise Model (Neural Network)
    \item $\theta$: Conditioning cosmology i.e. a vector with $\Omega_m$ and $\sigma_8$
\end{itemize}
\textsc{Computing VLB terms:}
\begin{align*}
& L_0 = - \log p_{\phi} (x_0|x_1, \theta) = -ln \mathcal{N}(x_0|\mu_0, \beta_0) = \sum_p \frac{(x_0 - \mu_0)^2}{2\beta_0} + 0.5ln|2\pi\beta_0|  \\*
& \textrm{For t} \in [1, T-1], L_t =  D_{KL}[q(x_{t}|x_{t+1}, x_0) || p_\phi (x_t|x_{t+1}, \theta)] \\*
& q(x_{t}|x_{t+1}, x_0) = \mathcal{N}(\tilde{\mu_t} (x_{t+1}, x_0), \tilde{\beta}_t) \\*
& \tilde{\mu_t} (x_{t+1}, x_0) = \frac{\sqrt{\bar{\alpha}_{t-1}\beta_{t}}}{1 - \bar{\alpha}_{t}} x_0 + \frac{\sqrt{\alpha_{t}(1 - \bar{\alpha}_{t-1})}}{1 - \bar{\alpha}_{t}} x_{t+1}  \text{ and } \tilde{\beta}_t = \beta_t \frac{1 - \bar{\alpha}_{t-1}}{1 - \bar{\alpha}_{t}}  \\*
& p_\phi (x_t|x_{t+1}, \theta) = \mathcal{N}(\tilde{\mu_t} (x_{t+1},  \hat{x}_{0, \phi}), \tilde{\beta}_t) \\*
& \hat{x}_{0, \phi} = \frac{x_{t+1} - \sqrt{1 - \bar{\alpha}_t \epsilon_\phi(x_{t+1}, t, \theta)}}{\sqrt{\bar{\alpha_t}}} \\*
\end{align*}

\subsection{HMC Convergence}
\label{app:hmc2}
\begin{figure*}
\centering
\includegraphics[width=0.45\linewidth]{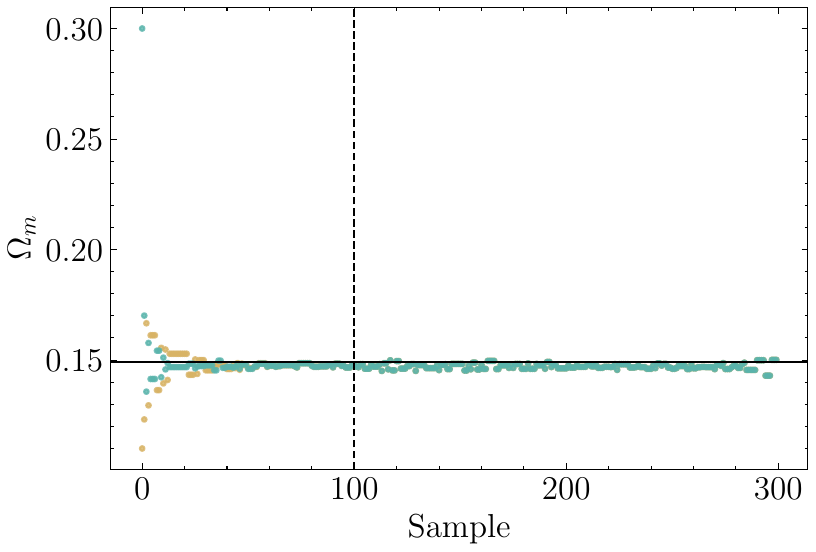}
\includegraphics[width=0.45\linewidth]{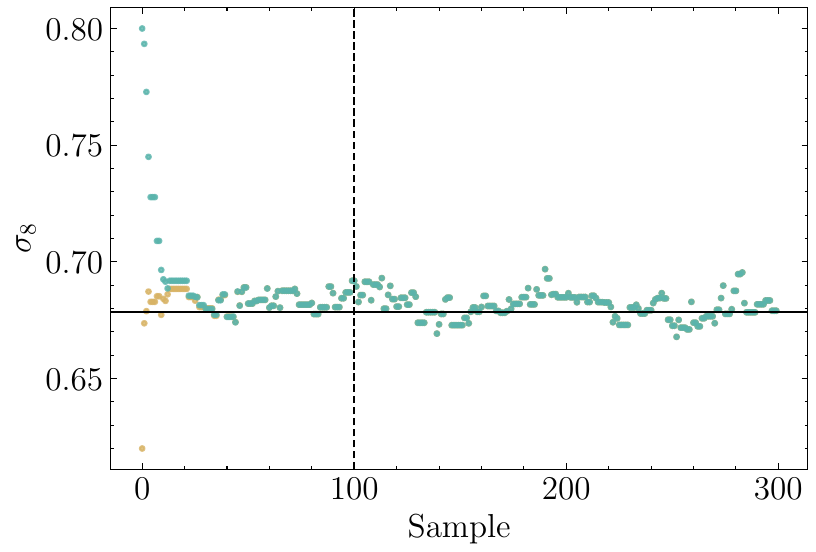}
\caption{HMC chains for 300 samples, for the same field, starting at two different initial parameters: [0.11, 0.62] (beige) and [0.3, 0.8] (teal). The chains are well mixed by the cutoff we designate as our burn-in (100 samples), denoted by the dashed line. }
\label{fig:hmc_rhat}
\end{figure*}

For a single field, we examine the convergence of parameters, when the chain starts from different initial parameters in Figure ~\ref{fig:hmc_rhat}. The chains are indistinguishable beyond around 50 samples. We thus choose a burn-in of 100 samples. The $\hat{R}$ for both parameters computed using the samples in $[100-300]$ is 0.997. A $\hat{R}$ of greater than 1.1 usually indicates that the chains have not converged and still retain some memory of their initialization. While the $\hat{R}$ is theoretically expected to be around 1 or slightly greater, some numerical variation about this expected value can result in values that are slightly less than 1. The $\hat{R}$ is a measure of the variance between chains divided by the variance within chains.

\subsection{Parameter Estimation Baselines}
\label{app:bl}
\textbf{Power Spectrum NPE Baseline:} We use a Masked Autoregressive Flow \citep{papamakarios2017masked} to implement the normalizing flow that predicts the 2 dimensional parameter vector given the 129 dimensional feature vector for a single field (128 bins for the power spectrum+1 for the mean of the log fields). The power spectrum is the log of the overdensity power spectrum of the (unlogged) fields. In Figure ~\ref{fig:hmc_main}, the power spectrum sample contours are smoothed by convolving with a Gaussian kernel with a scale of 0.8 and the Diffusion-HMC samples are smoothed by a kernel of 0.2. The ellipses in the inset figure are computed using the covariance of the 400 diffusion model samples and finding the ellipses corresponding to the 68.3, 99.4 and 99.7$\%$ confidence intervals, using the eigendecomposition of the covariance.

\subsection{Additional Robustness Tests}

\begin{figure*}
\centering
\includegraphics[width=0.9\linewidth]{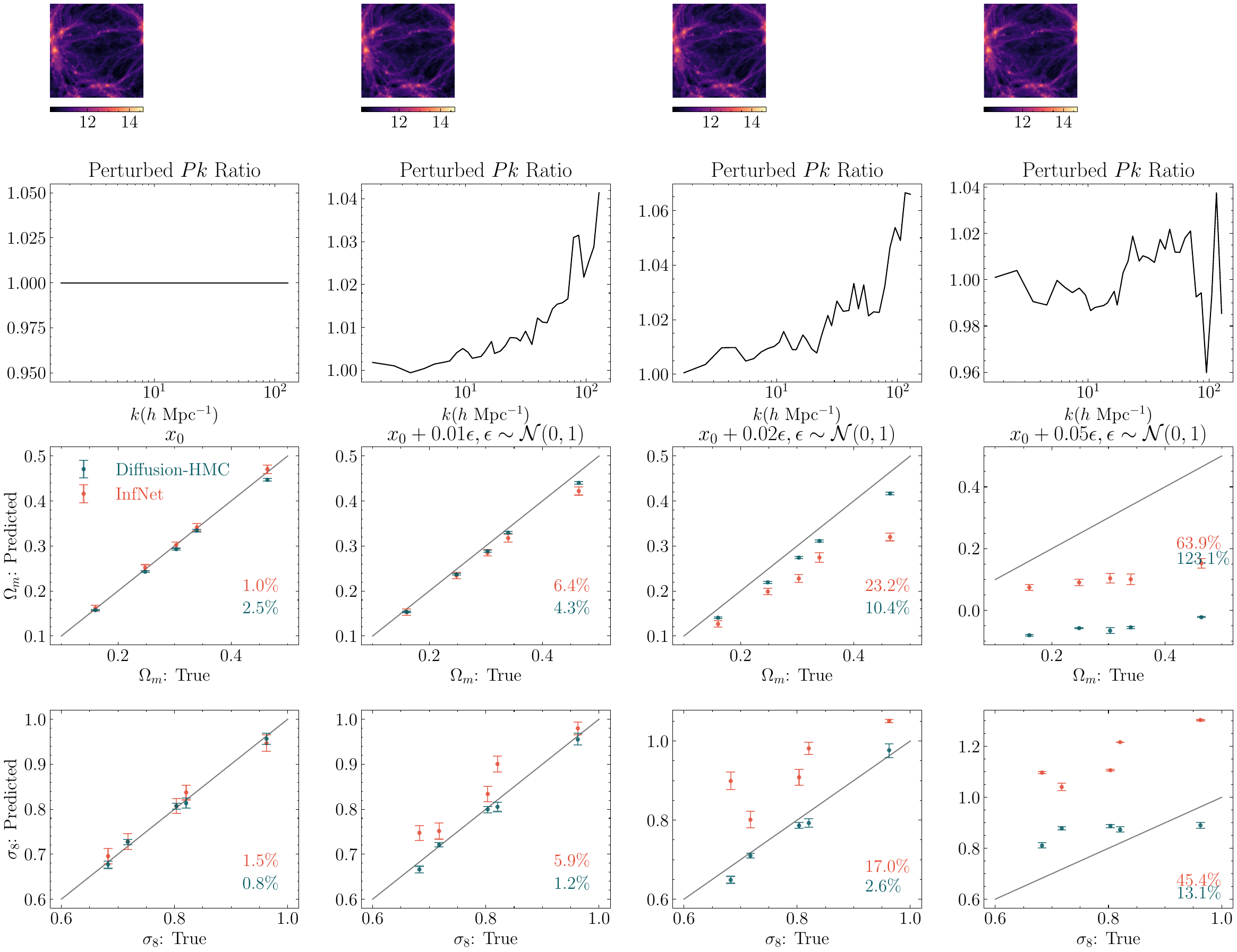}

\caption{Robustness comparison without a prior applied in the HMC case.}
\label{fig:robustness_noprior}
\end{figure*}

\paragraph{Dropping the Prior:} In Figure \ref{fig:robustness_noprior}, we explore robustness without the prior in the HMC setting, for the noise levels in Figure \ref{fig:robust} as well as with the addition of more noise. For small amounts of noise, ($\sigma$ = 0.01, 0.02) removing the prior doesn't affect the bias of the Diffusion-HMC-inferred parameters since the numbers in Figure 10 are identical / close to those in Figure \ref{fig:robust}. For noise with a scale of 0.05, the Diffusion-HMC constraints are more perturbed for $\Omega_m$ but more robust for $\sigma_8$ relative to the neural network baseline. A standard deviation of 0.05 is equivalent to the diffusion noising timesteps between the 1st and the 2nd timestep, while \{0.01, 0.02\} are less than the first timestep. Note, we plot the power spectra of the `true' \textit{(linear)} mass density field in these figures, to be consistent with the power spectra in Figures 1 and 2. However, the noise is added to the log of the field. Thus while the effect on the linear power spectrum is mild (around 5\% at the smallest scales) for $\sigma=0.05$ the effect on the power spectrum of the log field is around 35\% at the smallest scales.

\begin{figure*}
\centering
\includegraphics[width=0.32\linewidth]{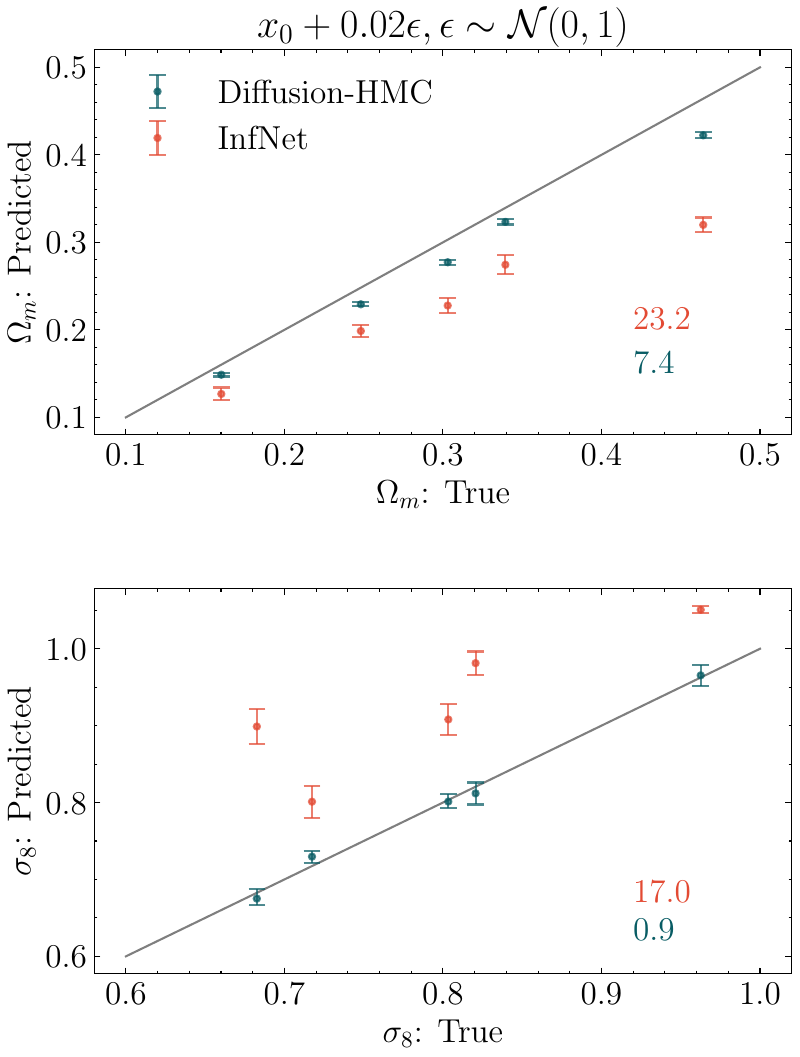}\includegraphics[width=0.32\linewidth]{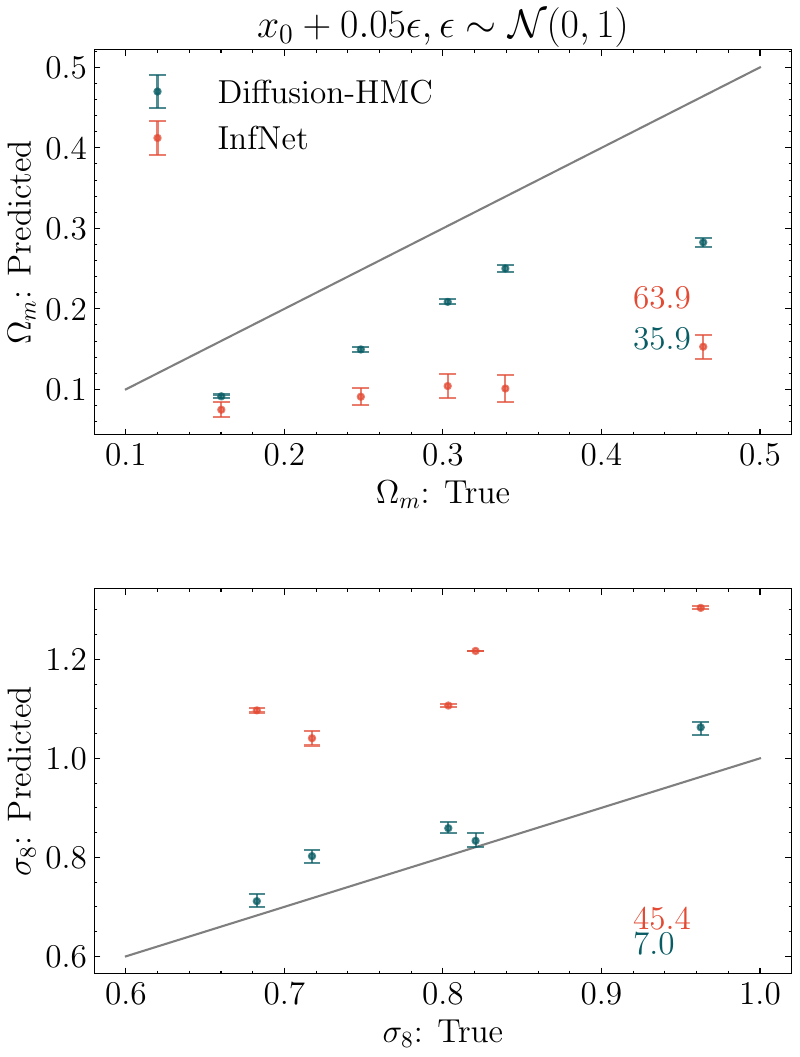}
\caption{Robustness comparison with the first 2 noising timesteps dropped \textit{and} without a prior applied in the HMC case.}
\label{fig:robustness_droppedt}
\end{figure*}

\paragraph{Dropping Initial Timesteps:} In the left panel of Figure \ref{fig:clik_t}, we observed that the isolated effect of the smallest timesteps also had the strongest constraints. We thus ask the question, if we perform parameter inference in a setting where there is some knowledge of the amount of noise added, can dropping timesteps confer additional robustness? In Figure \ref{fig:robustness_droppedt}, we add a comparison for the noising amounts corresponding to 0.02 and 0.05 and drop the first 2 timesteps. The rest of the setting is the same as in Figure \ref{fig:robustness_noprior} i.e. we also drop the prior. We find that dropping these timesteps confers greater robustness on the estimates, without noticeably reducing confidence. While we defer a more rigorous investigation to future work, the prospect of using the knowledge of the amount of noise added to strategically drop timesteps could be of interest, reminiscent of scale-dependent analyses and scale cuts in other cosmological analysis methods \citep{dai2024multiscale,krause2017dark}.

\section{Summary Statistics}
\subsection{Reduced Chi-squared Statistics}
\label{app:chi2}
\newcommand{\pref}{Pk_{Ref}}
\newcommand{\ptest}{Pk_{Test}}
For the CV fields, where we have 450 samples of the true and generated fields for a single parameter, we compute the reduced chi-squared statistic using an estimate of the covariance between different $k$ bins. The number of $k$ bins here is 35.

\begin{align}
& \vec{\mu} = \langle \pref \rangle \quad\quad\quad  C = \rm{Cov}[\pref] \quad\quad\quad \hat{C}^{-1} = C^{-1} \frac{N-p-2}{N - 1} \label{eq:hartlap} \\*
& \chi^2_r(\ptest)= \sum_k (\ptest - \vec{\mu})\cdot (\hat{C}^{-1} (\ptest - \vec{\mu})^T)^T  \label{eq:multidim_chisq}
\end{align}

For the LH fields during model selection, since we just have 15 fields in the true dataset, we cannot reliably estimate a covariance. We use the following formula instead. The number of $k$ bins here is 128.
\begin{align}
\chi_r^2 (s) = \frac{1}{|k|}\sum_k\frac{(P(k)_s - <P(k)_{\textsc{True}}>)^2}{\sigma[P(k)_\textsc{True}]^2} \label{eq:unichisq}
\end{align}

\subsection{Across Checkpoints}
For the eight checkpoints we generated 500 fields, with 50 fields for each of the 10 validation parameters. We then examined how different the reduced chi-squared statistic of the power spectra of the log fields, the linear fields and the $p$ values of the means of the distributions of true and generated fields. A value of less than 0.05 would indicate that the two distributions of the means are statistically different. The $p$ values are above 0.05 for all eight checkpoints. Since these comparisons are limited by the number of samples in the true set for each parameter (15), we additionally plot the reduced chi-squared statistic derived by using each true fields as the test and the other 14 fields as the reference. While there is some oscillation across checkpoints for each of these three statistics, the variation appears random.

\begin{figure*}[h!]
\centering
\includegraphics[width=0.32\linewidth]{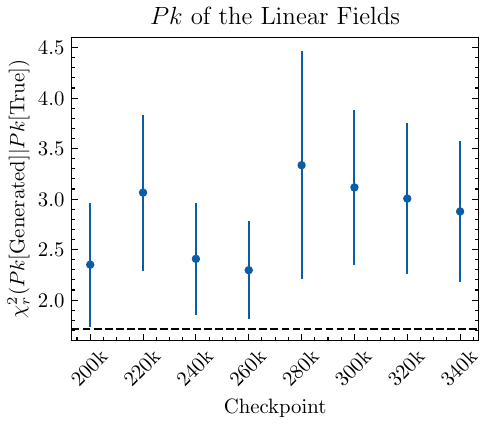}
\includegraphics[width=0.32\linewidth]{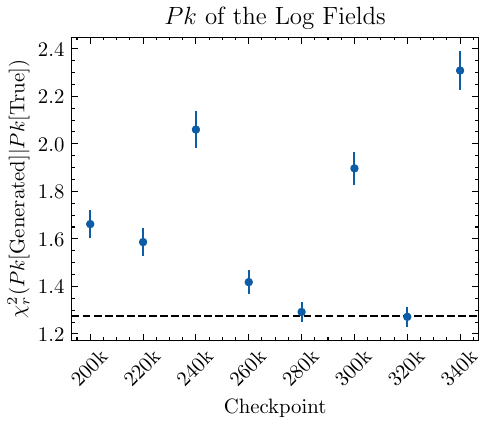}
\includegraphics[width=0.32\linewidth]{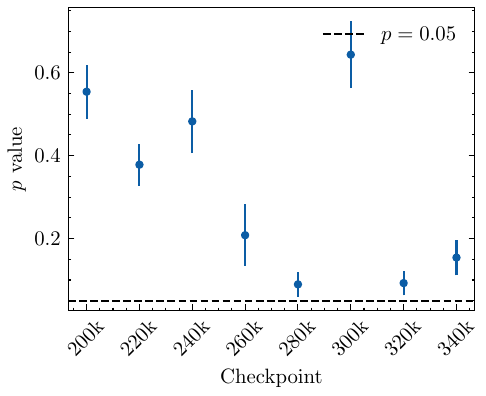}
\caption{Mean and standard error on the mean for the reduced $\chi_r^2$ statistic of the power spectra of the 50 linear (\textbf{left}) and log (\textbf{center}) fields for each parameter relative to the 15 true fields' power spectra for that parameter, across 10 different parameters for each of 8 checkpoints. The dashed line demarcates the mean reduced $\chi_r^2$ of the true fields using the other 14 true fields as the reference distribution (leave-one-out). \textbf{Right:} Mean and standard error on the mean of the $p$ values of the distribution of the means of the 50 generated log fields relative to the distribution of the means of the 15 true fields for the same parameter. The $p$ values are above 0.05 for all of the 8 checkpoints. }
\label{fig:ckpwise}
\end{figure*}

\end{document}